\documentclass[prl,twocolumn,amsmath,amssymb,amsfonts,final,letterpaper,floatfix]{revtex4}

\usepackage{graphicx}
\usepackage{hyperref}
\newcommand{\dof}{DOF}
\newcommand{\oam}{OAM}

\newcommand\bsa{BSA}
\newcommand\hbsa{HBSA}
\newcommand\poam{$\pm1$-\oam}
\newcommand\oaml{\ensuremath{\circlearrowleft}}
\newcommand\oamr{\ensuremath{\circlearrowright}}

\begin{document}

\title{Beating the channel capacity limit for linear photonic superdense
coding}

\author{Julio T. Barreiro}
\author{Tzu-Chieh Wei$^*$}
\author{Paul G. Kwiat}
\affiliation{Department of Physics, University of Illinois at %
Urbana-Champaign, Urbana, Illinios 61801-3080, USA\\
$^*$ Present address: Institute for Quantum Computing and Department of
Physics and Astronomy, University of Waterloo,
Waterloo, ON N2L 3G1, Canada.}

\begin{abstract}
Dense coding is arguably the protocol that launched the field of quantum
communication~\cite{bennett-prl-69-2881}.  Today, however, more than a decade
after its initial experimental realization~\cite{mattle-prl-76-4656}, the
channel capacity remains fundamentally limited as conceived for photons using
linear elements.  Bob can only send to Alice three of four potential messages
due to the impossibility of performing the deterministic discrimination of all
four Bell states with linear
optics~\cite{vaidman-pra-59-116,lutkenhaus-pra-59-3295}, reducing the
attainable channel capacity from 2 to $\log_23\approx1.585$ bits.  However,
entanglement in an extra degree of freedom enables the complete and
deterministic discrimination of all Bell
states~\cite{kwiat-pra-58-R2623,schuck-prl-96-190501,barbieri-pra-75-042317}.
Using pairs of photons simultaneously entangled in spin and orbital angular
momentum~\cite{molina-terriza-nphys-3-305,barreiro-prl-95-260501}, we
demonstrate the quantum advantage of the ancillary entanglement.  In
particular, we describe a dense-coding experiment with the largest reported
channel capacity and, to our knowledge, the first to break the conventional
linear-optics threshold.  Our encoding is suited for quantum communication
without alignment~\cite{aolita-prl-98-100501} and satellite communication.
\end{abstract}

\maketitle 

The first realization of quantum dense coding was optical, using pairs of
photons entangled in polarization~\cite{mattle-prl-76-4656}.  Dense coding
has since been realized in various physical systems and broadened
theoretically to include high-dimension quantum states with
multiparties~\cite{liu-pra-65-022304}, and even coding of quantum
states~\cite{harrow-prl-92-187901}.  The protocol extension to continuous
variables~\cite{braunstein-pra-61-042302,ban-qso-1-l9} has also been
experimentally explored optically, using superimposed squeezed
beams~\cite{li-prl-88-047904}.  Other physical approaches include a
simulation in nuclear magnetic resonance with temporal
averaging~\cite{fang-pra-61-022307}, and an implementation with atomic qubits
on demand without postselection~\cite{schaetz-prl-93-040505}.  However,
photons remain the optimal carriers of information given their resilience to
decoherence and ease of creation and transportation.

Quantum dense coding was conceived~\cite{bennett-prl-69-2881} such that Bob
could communicate two bits of classical information to Alice with the
transmission of a single qubit, as follows.  Initially, each party holds one
spin-$\frac{1}{2}$ particle of a maximally entangled pair, such as one of the
four Bell states.  Bob then encodes his 2-bit message by applying one of four
unitary operations on his particle, which he then transmits to Alice.
Finally, Alice decodes the 2-bit message by discriminating the Bell state of
the pair.

Alice's decoding step, deterministically resolving the four Bell states, is
known as Bell-state analysis (\bsa{}).  While in principle attainable with
nonlinear interactions, such \bsa{} with photons is very difficult to achieve
with present technology, yielding extremely low efficiencies and low
discrimination fidelities~\cite{kim-prl-86-1370}.  Therefore, current
fundamental studies and technological developments demand the use of linear
optics.  However, for quantum communication, standard \bsa{} with linear optics
is fundamentally impossible~\cite{vaidman-pra-59-116,lutkenhaus-pra-59-3295}.
At best only two Bell states can be discriminated; for quantum communication
the other two are considered together for a three-message encoding.
Consequently, the maximum channel capacity of this conventional optical dense
coding is $\log_2 3\approx1.585$ bits.  Although there are probabilistic
approaches that can distinguish all 4 Bell states (which would be necessary to
achieve the fundamental channel capacity of 2), these are at best successful
50\% of the time~\cite{calsamiglia-apb-B72-67}, so have a net channel capacity
of at most 1 per photon.

Entanglement in an extra degree of freedom (\dof{}) of the pair,
hyperentanglement~\cite{kwiat-jmo-44-2173}, enables full \bsa{} with linear
optics~\cite{kwiat-pra-58-R2623,schuck-prl-96-190501}.  In this case, since Bob
only encodes information in one \dof{} (the auxiliary \dof{} is unchanged), a
dense-coding protocol proceeds under the same encoding conditions as in the
original proposal~\cite{bennett-prl-69-2881}.  Although
hyperentanglement-assisted \bsa{} (\hbsa{}) on polarization states has been
reported with ancillas entangled in energy-time~\cite{schuck-prl-96-190501} and
linear-momentum~\cite{barbieri-pra-75-042317}, no advantage for quantum
information or fundamental physics was shown; experiments thus far have been
limited to a channel capacity of less than 1.18(3)
bits~\cite{schuck-prl-96-190501}, substantially less than is possible even
without hyperentangled resources.

\begin{figure*}
\centerline{\includegraphics{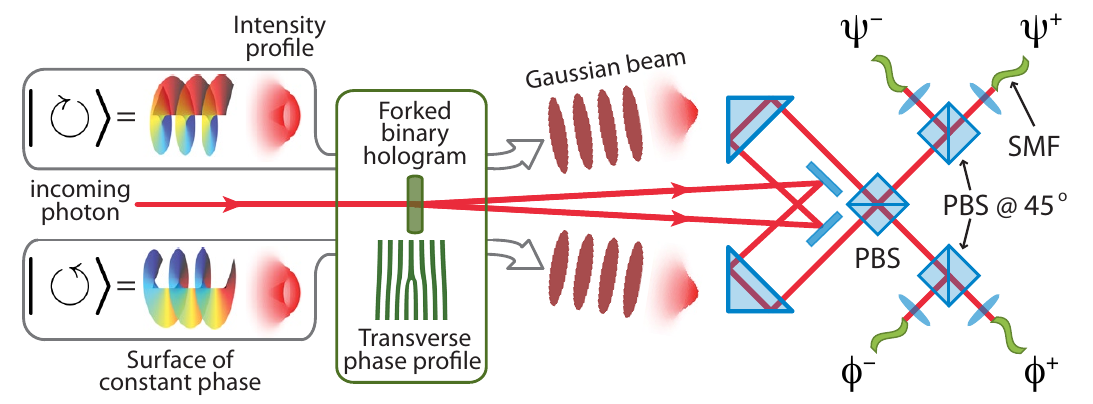}}
\caption{\textbf{Spin-orbit Bell-state analyzer.}  A photon in a spin-orbit
  Bell-state incident from the left is first split according to its \poam{}
  content; its \poam{} components are converted to 0-\oam{} and combined on a
  polarizing beam splitter (PBS) for a spin-controlled orbit-CNOT gate.  The
  photon is then filtered by a single mode fibre (SMF) and finally routed to a
  unique detector (photon-counting avalanche photodiode).}
\label{fig:splitter}
\end{figure*}

Using pairs of photons entangled in their spin and orbital angular momentum
(\oam{}) in a \hbsa{} with high stability and high detection fidelity, we
realize a dense-coding experiment with a channel capacity that exceeds the
threshold to beat conventional linear-optics schemes.  In our scheme, Alice and
Bob are provided with pairs of photons simultaneously entangled in their spin
and \poam{} in a state of the form
\begin{equation}
\frac{1}{2}(|HH\rangle+|VV\rangle)\otimes(|\oaml\oamr\rangle+|\oamr\oaml\rangle).
\label{eq:epr}
\end{equation}
Here $H$ ($V$) represents the horizontal (vertical) photon polarization and
$\oaml$ ($\oamr$) represents the paraxial spatial modes (Laguerre-Gauss)
carrying $+\hbar$ ($-\hbar$) units of \oam{}~\cite{allen-2003-oam}.  Bob
encodes his message by applying one of four unitary operations on the spin of
his photon of this hyperentangled pair: (1) the identity, (2) $V\rightarrow-V$,
(3) $H\leftrightarrow V$, or (4) $V\rightarrow-V$ and $H\leftrightarrow V$.
Such operations transform the state in equation~\ref{eq:epr} into
\begin{eqnarray}
\Phi^\pm_\text{spin}\otimes\Psi^+_\text{orbit},\,\,\text{and}\,\,
\Psi^\pm_\text{spin}\otimes\Psi^+_\text{orbit},
\label{eq:encoded}
\end{eqnarray}
where the spin and orbit Bell-states are defined as
\begin{eqnarray*}
\Phi^\pm_\text{spin}&\equiv&\left(|HH\rangle\pm|VV\rangle\right)/\sqrt{2},\\
\Psi^\pm_\text{spin}&\equiv&\left(|HV\rangle\pm|VH\rangle\right)/\sqrt{2},\\
\Psi^+_\text{orbit}&\equiv&\left(|\oaml\oamr\rangle+|\oamr\oaml\rangle\right)/\sqrt{2}.
\end{eqnarray*}

We designed a \hbsa{} scheme (inspired by Ref.~\cite{walborn-pra-68-042313})
enabling Alice to discriminate the four states in equation ~\ref{eq:encoded}.
In this scheme, the polarization \bsa{} relies on the observation that the
states resulting from Bob's encoding can be rewritten as superpositions of
the single-photon Bell-states of spin and orbital angular momentum, or
spin-orbit Bell-states:
\begin{eqnarray*}
\phi^\pm &\equiv&
\frac{1}{\sqrt{2}}(|H\oaml\rangle\pm|V\oamr\rangle),\\ \psi^\pm
&\equiv& \frac{1}{\sqrt{2}}(|H\oamr\rangle\pm|V\oaml\rangle).
\end{eqnarray*}
In this basis, the states Alice analyzes have the form
\begin{align*}
\Phi^\pm_\text{spin}\otimes\Psi^+_\text{orbit} =
\frac{1}{2}\big(&\phi^+_1\otimes\psi^\pm_2 + \phi^-_1\otimes\psi^\mp_2
\\ & + \psi^+_1\otimes\phi^\pm_2 + \psi^-_1\otimes\phi^\mp_2\big),
\\ \Psi^\pm_\text{spin}\otimes\Psi^+_\text{orbit} =
\frac{1}{2}\big(&\pm\phi^+_1\otimes\phi^\pm_2 \mp \phi^-_1\otimes\phi^\mp_2
\\ &\pm \psi^+_1\otimes\psi^\pm_2 \mp \psi^-_1\otimes\psi^\mp_2\big).
\end{align*}
This arrangement shows that each hyperentangled state is a unique
superposition of four of the sixteen possible combinations of 2-photon
spin-orbit Bell-states.  Therefore, Alice can decode Bob's message by
performing spin-orbit \bsa{} locally on each photon.

We implement the spin-orbit \bsa{} with a novel interferometric apparatus
consisting of a \poam{} splitter and polarizing beam splitters (PBS), as shown
in Fig.~\ref{fig:splitter}.  The first splitter combines the action of a binary
plane-wave phase grating~\cite{allen-2003-oam} and single-mode fibres.  The
grating transforms an incoming photon in the state $|\!\oaml\rangle$
($|\!\oamr\rangle$) into a gaussian beam with no \oam{} in the $+1$ ($-1$)
diffraction order (for a splitter that preserves the photon's \oam{}, see
Ref.~\cite{allen-2003-oam}).  Subsequently filtering the first diffraction
orders with single-mode fibres, we effectively split an incoming photon into
its \poam{} components.  By merging these diffraction orders on a PBS we
perform a spin-controlled NOT gate over the photon \oam.  In
Fig.~\ref{fig:splitter}, the states $\psi^\pm$ ($\phi^\pm$) exit on the top
(bottom) output port of the PBS.  Followed by measurements in the diagonal
basis, shown in Fig.~\ref{fig:splitter} as PBS@$45^\circ$, the desired
measurement in the single-photon Bell-state basis is accomplished.  Additional
birefringent elements make this device a universal unitary gate for
single-photon two-qubit states, in analogy with the device for
polarization-linear momentum states in Ref.~\cite{englert-pra-63-032303}.

\begin{figure}[h!]
\centerline{\includegraphics{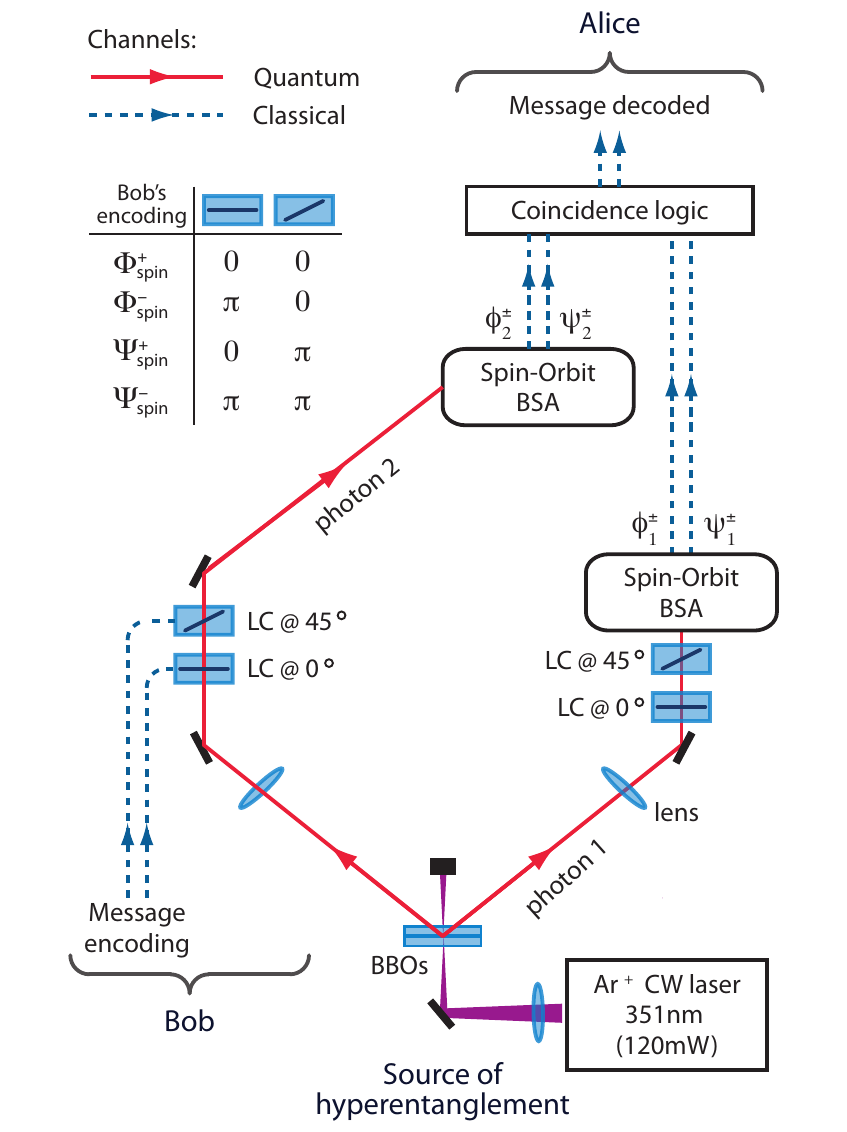}}
\caption{\textbf{Experimental setup for dense coding with spin-orbit encoded
    photons.}  Acting on photon 2 of a hyperentangled pair, Bob encodes his
  message by using the liquid crystals (LCs) to apply the phases indicated in
  the table, while (or earlier) Alice performs spin-orbit \bsa{} on photon 1.
  Later --the upward direction suggests time progression-- Alice uses a
  spin-orbit \bsa{} on photon 2, and the result from the measurement on photon
  1, to decode Bob's message.  The liquid crystals on the path of photon 1
  applied no phase during the dense-coding experiment, but were used along with
  Bob's liquid crystals to characterize the polarization states of the
  hyperentangled source by quantum state tomography.  The liquid-crystal optic
  axes are perpendicular to the incident beams; LC@$45^\circ$ (LC@$0^\circ$) is
  oriented at $45^\circ$ ($0^\circ$) from the horizontal polarization
  direction. BBOs: $\beta$-barium borate nonlinear crystals; CW:
  continuous-wave.}
\label{fig:setup}
\end{figure}

Each step in the dense-coding protocol corresponds to a distinct experimental
stage in Fig.~\ref{fig:setup}: a hyperentanglement source, Bob's encoding
components, and Alice's \hbsa{}.  The hyperentanglement source is realized via
spontaneous parametric downconversion in a pair of nonlinear crystals (see
Methods section).  The generated photon pairs are entangled in polarization,
\oam{} and emission time~\cite{barreiro-prl-95-260501}.  In particular, we use
a subspace of the produced states which was shown to have a state overlap or
fidelity of 97\% with the state in equation~\ref{eq:epr}.  Next, Bob encodes
his message in the polarization state by applying birefringent phase shifts
with a pair of liquid crystals, as shown in Fig.~\ref{fig:setup}.  Finally,
Alice performs \hbsa{} using two of the spin-orbit Bell-state analyzers shown in
Fig.~\ref{fig:splitter}, one for each photon (see Methods section).

\begin{figure*}
\centerline{\includegraphics{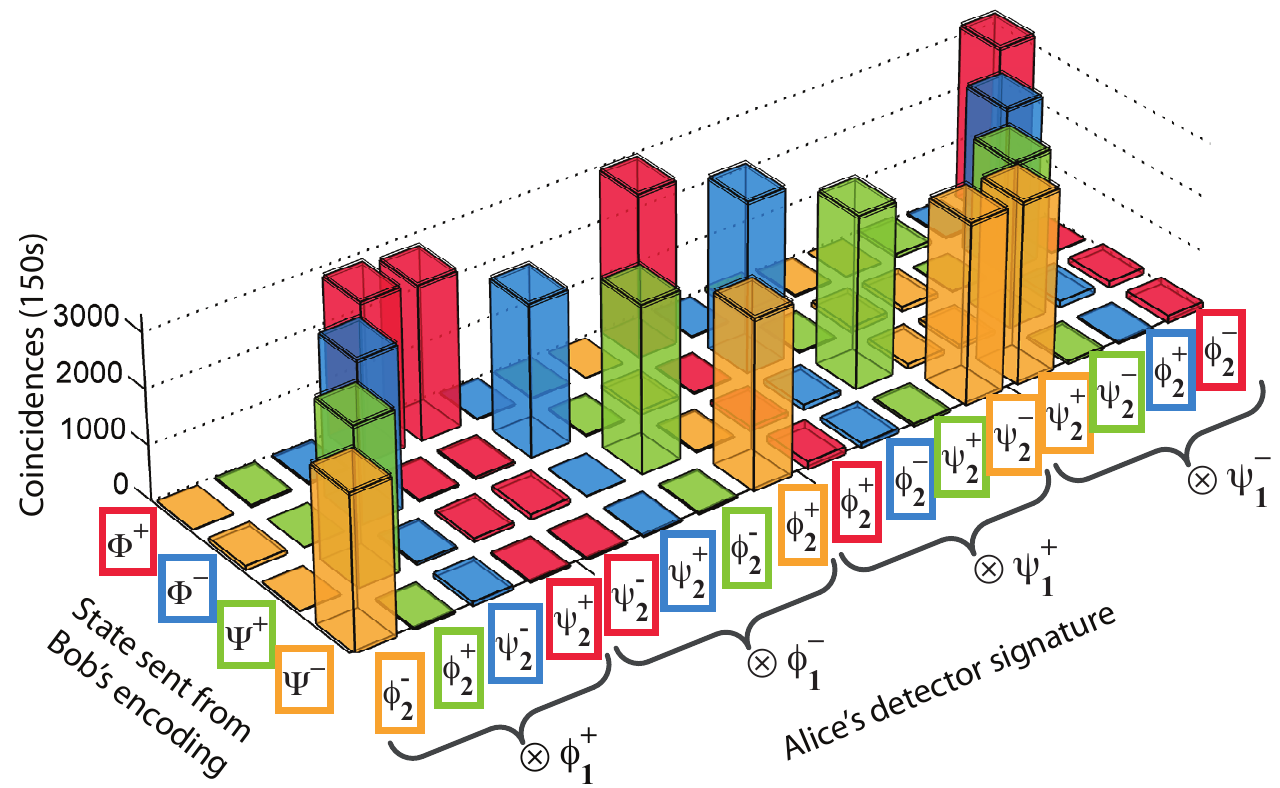}}
\caption{\textbf{Experimental results of hyperentanglement-assisted dense
    coding.}  Coincidence counts detected by Alice's \hbsa{} for each message
  (state) sent by Bob.  The error bars (shown as additional squares at the top
  of each column) represent $\pm1$ standard deviations, deduced from poissonian
  counting statistics.  The state-discrimination signal-to-noise ratios (SNR),
  which compare the sum of the four rates corresponding to the actual state to
  the sum of the other twelve registered rates, are $SNR_{\Phi^+}=19.9(8)$,
  $SNR_{\Phi^-}=27(1)$, $SNR_{\Psi^+}=13.7(5)$, and $SNR_{\Psi^-}=16.4(6)$.}
\label{fig:counts}
\end{figure*}

\begin{figure}
\centerline{\includegraphics{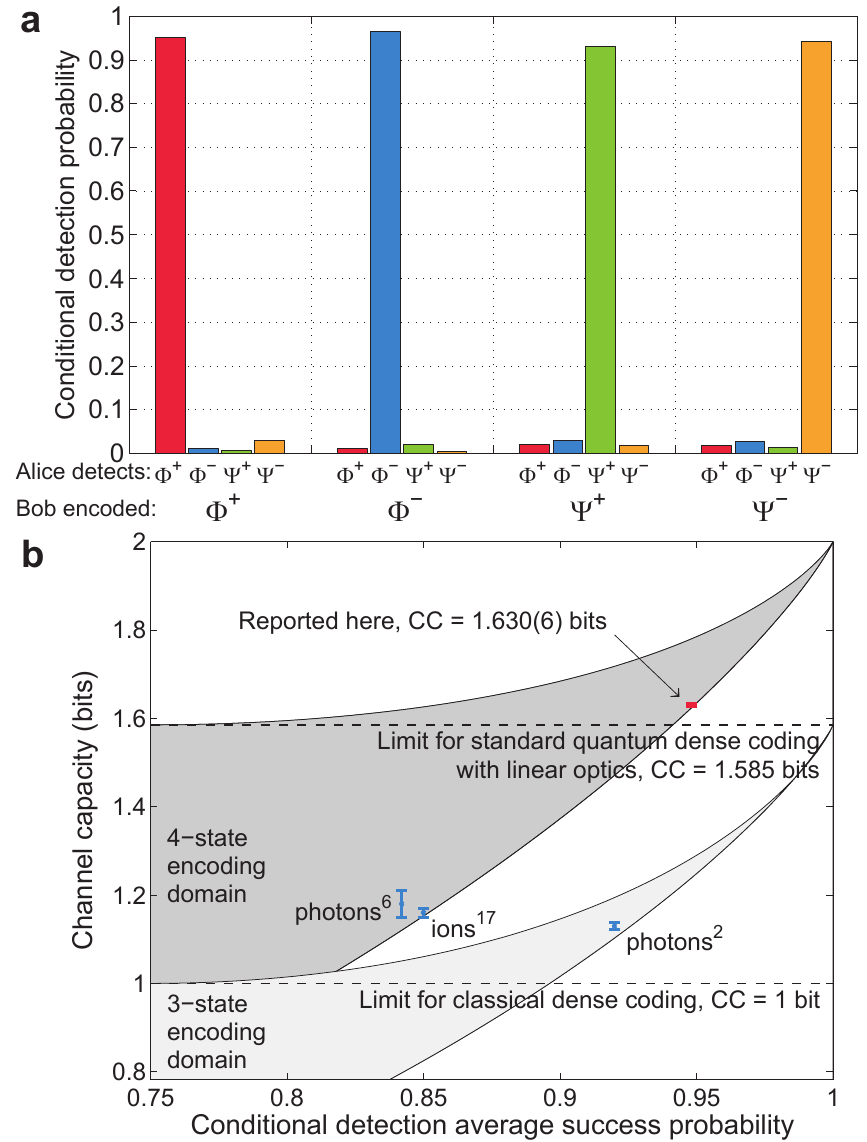}}
\caption{\textbf{Conditional detection probabilities beating the channel
    capacity limit for standard dense-coding with linear optics.} \textbf{a}
  Given that Bob encoded the four states indicated, Alice infers the state
  transmitted with the probabilites shown (calculated from data in
  Fig.~\ref{fig:counts}).  Her average success probability is $94.8(2)\%$.  The
  uncertainty in each probability is less than 0.2\%.  These results imply a
  channel capacity of 1.630(6) bits, above the standard linear-optics limit of
  1.585. \textbf{b} Experimentally reported channel capacities as a function of
  their conditional detection average success probability.  The error bars
  represent the statistical error of $\pm1$ standard deviations.  The domains
  of achievable channel capacity for both 3- and 4-state encodings are shown
  for reference (see Supplementary Information I).}
\label{fig:probs}
\end{figure}

We characterize our dense-coding implementation by switching between the four
states for equal intervals, and measuring all output states of the \hbsa{}.
The result of these measurements are coincidence counts for each input state,
as shown in Fig.~\ref{fig:counts}.  From this data we can determine the
conditional detection probabilities that Alice detects each message
$\Phi^\pm$ and $\Psi^\pm$ given that Bob sent, for example, the message
$\Phi^+$.  The probabilities shown in Fig.~\ref{fig:probs} were calculated by
comparing the sum of the four rates corresponding to each detected message
over the sum of all sixteen rates for the sent message.  The average
probability of success was 94.8(2)\% (all reported errors from Monte Carlo
simulations).

A better figure of merit for a quantum dense-coding implementation is the
channel capacity, since it characterizes the exponential growth of the maximum
number of distinguishable signals for a given number of uses of the channel
(see Methods section).  From the conditional detection probabilities, we obtain
a channel capacity of 1.630(6) bits with a probability of sending each state of
$P(\Phi^+)=0.26$, $P(\Phi^-)=0.26$, $P(\Psi^+)=0.24$, and $P(\Psi^-)=0.24$.
This exceeds the 1.585 channel-capacity threshold for conventional
linear-optics implementations.  The channel capacity drifted by no more than
one standard deviation between experimental runs, demonstrating the high
stability of the implementation.

The experimental channel capacity is nevertheless smaller than the maximum
attainable (2 bits), due to imperfections in the alignment, input states and
components.  By characterizing each imperfection and modelling the gates and
measurement, we estimated their effect on the channel capacity (see
Supplementary Information II).  Considering all mentioned imperfections (see
Methods section) and their spread in a Monte Carlo simulation, the predicted
channel capacity of 1.64(2) bits agrees with the measured channel capacity of
1.630(6) bits.  The polarization and spatial-mode states can be improved by
spatially compensating the angle-dependent phase~\cite{altepeter-oe-13-8951},
using a forked hologram with a smaller diffraction angle to decrease wavelength
dispersion (a potential source of alignment imbalances), and obtaining crystals
with a smaller wedge.  The deleterious effect of the PBS crosstalk can be
reduced by adding extra phase-compensation plates inside the interferometers,
and can potentially be eliminated altogether by adding appropriate birefringent
beam displacers after each PBS.

Above, Bob encoded two qubits in the form of spin-orbit Bell-states by acting
only on the spin \dof{}.  However, more generally he could also apply one of
four unitaries in the \poam{} subspace and encode four qubits.  The state of
the pair of photons then becomes a product of Bell states, 16 in total.  In
principle, if Alice could discriminate all these ``hyper-Bell'' states, up to 4
bits could be transmitted per photon.  We have investigated the limits for
unambiguously distinguishing these Bell-like states, and have found that the
optimal one-shot discrimination scheme is to group the 16 states into 7
distinguishable classes~\cite{wei-pra-2007}. The optimal analysis can be
achieved by the Kwiat-Weinfurter scheme~\cite{kwiat-pra-58-R2623}, with
photon-number resolving detectors, giving a maximum channel capacity of $\log_2
7\approx 2.81$ bits.  If we modify the present scheme, we can also implement an
unambiguous discrimination of all 16 Bell states with two identical
copies~\cite{wei-pra-2007}.

In conclusion, we have beaten a fundamental limit on the channel capacity for
standard dense coding using only linear optics.  A number of features make our
\hbsa{} efficient and reliable.  First, hyperentanglement offers advantages in
the source, logic gates and detection side.  Quantum logic between qubits
encoded on different \dof{}s is much more easily implemented than when using
different photons~\cite{cerf-pra-57-R1477,fiorentino-prl-93-070502}.  From the
source side, more quantum information is available per photon, particularly
with the energy-time and spatial-mode \dof{}(e.g.,
\cite{ali-khan-prl-98-060503}).  In the detection side, compared to
multi-photon approaches, higher efficiency is achieved because only one pair of
photons is detected.  Second, since our \hbsa{} requires only local
measurements, Alice can measure one of the photons and store the classical
result of her measurement until Bob sends his photon (i.e., she does not
require a quantum memory).  Finally, the photon's polarization and \poam{}
constitute a robust encoding as they enable quantum communication without
alignment~\cite{aolita-prl-98-100501} as well as other landmark advances for
quantum information~\cite{molina-terriza-nphys-3-305}.  Furthermore, by using
paraxial beams as the ancillary \dof{}, the scheme is free of tight
source-to-detector requirements such as interferometric
stability~\cite{barbieri-pra-75-042317} or perfect indistiguishability for HOM
interference~\cite{schuck-prl-96-190501}.  However, \oam{} single-photon and
entangled states easily decohere by atmospheric
turbulence~\cite{paterson-prl-94-153901,smith-pra-74-062104}, limiting their
likely communication applications to satellite-to-satellite transmissions.

\section{Methods}

The hyperentanglement source is realized by directing 120 mW of 351 nm light
from a continuous-wave (CW) Ar$^+$ laser into two contiguous $\beta$-barium
borate (BBO) nonlinear crystals with optic axes aligned in perpendicular
planes~\cite{barreiro-prl-95-260501}.  Type-I degenerate 702 nm photons in a
3.0$^\circ$ half-opening angle cone are produced by phase-matching each
0.6-mm-thick crystal.  In the spin and $\pm1$ \oam{} subspace, a two-fold
coincidence rate of 5 detected pairs/s is determined by a 10 ns coincidence
window and interference filters with $\Delta\lambda_\text{FWHM}=5\;$nm.

In our \hbsa{} implementation each PBS@$45^\circ$ and its two outputs in the
spin-orbit \bsa{} (Fig.~\ref{fig:splitter}) were replaced by a dichroic
polarizer oriented at either $45^\circ$ or $-45^\circ$ and a single output;
Alice's \hbsa{} thus acquires all spin-orbit \bsa{} outputs from four polarizer
settings.  With the CW source, Alice cycles through the four polarizer
settings, and for each polarizer setting Bob encodes the four messages, each
for 150 seconds.  During the measurement, no active stabilization or
realignment was done on the source, spin-orbit \bsa{} interferometers, or
coupling optics.  The \hbsa{} polarizers and liquid crystals were quickly set
with computer-controlled rotation stages and liquid crystal controllers.

The wavelength-dependent voltage applied to each liquid crystal was
independently calibrated to produce a birefringent phase difference of 0 or
$\pi$ with a diode laser operated at 699 nm (Hitachi HL-6738MG, driven at 140mA
and $80^\circ$C); the same laser was used to align the \poam{} splitter.  The
binary forked holograms were silver-halide emulsion gratings with
33\%-diffraction efficiency into the first order (more efficient schemes are
described in Ref.~\cite{allen-2003-oam}).  The same holographic plate included
spatial-mode tomography patterns, which in conjunction with the liquid crystals
were used for state reconstruction~\cite{barreiro-prl-95-260501}.  The spurious
phase upon reflection on the PBS was compensated with a waveplate in each
output port of the PBS for both spin-orbit Bell-state analyzers.  The state
discrimination signal-to-noise ration (SNR) varied between states due to
mode-coupling imbalance in the spin-orbit \bsa{}, PBS crosstalk, and slight
offsets in the liquid crystal calibrations.

We characterized the source polarization state $\Phi^-_\text{spin}$ by quantum
state tomography in the $|\oaml\oamr\rangle$ and $|\oamr\oaml\rangle$ \oam{}
subspaces~\cite{barreiro-prl-95-260501} (using the liquid crystals shown in
Fig.~\ref{fig:setup} and PBS of each spin-orbit \bsa{} shown in
Fig.~\ref{fig:splitter}).  Considering all combinations of signature detectors,
we measured an average degree of entanglement or tangle of $T=96.7(8)\%$ and a
mixture or linear entropy of $S_L=2.0(4)\%$.  If such high-quality polarization
state were exactly the same for each combination of signature detectors, the
decrease in the channel capacity would only be 0.006 bits.  However, small
differences in the coupled state between each combination of detectors
(expressed above as uncertainty) result in a channel capacity decrease of
0.09(2) (see Supplementary Information II).  The \oam{} state was also
tomographically reconstructed in the $|HH\rangle$ and $|VV\rangle$ polarization
subspaces~\cite{barreiro-prl-95-260501}, measuring an average $T=91(3)\%$ and
$S_L=6(2)\%$, yielding a channel capacity decrease of 0.20(3) bits.  The PBS
crosstalk (0.5\% for $H$, 1.0\% for $V$) further decreases the channel capacity
by 0.10(1) bits.  Finally, accidental coincidences (5 in 150s) reduce channel
capacity by 0.02 bits.

\subsection*{Channel capacity}

The capacity of a noisy channel is given by $\max_{p(x)} H(X:Y)$, where $x$ is
in the space of signals that can be transmitted $X$, $H(X:Y)$ is the mutual
information of $X$ and the space of received signals $Y$, and the maximum is
taken over all input distributions $p(x)$.  $H(X:Y)$ is a function of $p(x)$
and the conditional detection distribution $p(y|x)$ of receiving $y$ given that
$x$ was sent:
$$ H(X:Y) = \sum_{y \in Y} \sum_{x \in X} p(x) p(y|x)
\log{\frac{p(y|x)}{\sum_{x \in X} p(y|x) p(x)}}.
$$ In our experiment, a uniform probability of transmission gives a mutual
information of 1.629(6) bits, negligibly smaller than the channel capacity due
to the nearly balanced conditional probabilities, i.e., there is little to be
gained by sending some states more frequently.

Correspondence and requests for materials should be addressed to
J.T.B. e-mail: julio.barreiro@gmail.com

\section{Acknowledgements}

We thank N. Peters and N. Langford for helpful discussions and R. Hillmer for
assistance in hologram fabrication.  This work was jointly supported by the
DTO/ARO-sponsored MURI Center for 
Photonic Quantum Information Systems, and
the DTO/IARPA-sponsored Advanced Quantum Communication grant.
J.T.B. acknowledges support from CONACYT-M\'{e}xico.

Supplementary Information accompanies this paper.

\section{Competing financial interests}
The authors declare no competing financial interests.

\clearpage

\begin{widetext}

\end{widetext}


\begin{thebibliography}{30}
\expandafter\ifx\csname url\endcsname\relax
  \def\url#1{\texttt{#1}}\fi
\expandafter\ifx\csname urlprefix\endcsname\relax\def\urlprefix{URL }\fi
\providecommand{\bibinfo}[2]{#2}
\providecommand{\eprint}[2][]{\url{#2}}

\bibitem{bennett-prl-69-2881}
\bibinfo{author}{Bennett, C.~H.} \& \bibinfo{author}{Wiesner, S.~J.}
\newblock \bibinfo{title}{Communication via one- and two- particle operators on
  {E}instein-{P}odolsky-{R}osen states}.
\newblock \emph{\bibinfo{journal}{Phys. Rev. Lett.}}
  \textbf{\bibinfo{volume}{69,}} \bibinfo{pages}{2881--2884}
  (\bibinfo{year}{1992}).

\bibitem{mattle-prl-76-4656}
\bibinfo{author}{Mattle, K.} \emph{et~al.}
\newblock \bibinfo{title}{Dense coding in experimental quantum communication}.
\newblock \emph{\bibinfo{journal}{Phys. Rev. Lett.}}
  \textbf{\bibinfo{volume}{76,}} \bibinfo{pages}{4656--4659}
  (\bibinfo{year}{1996}).

\bibitem{vaidman-pra-59-116}
\bibinfo{author}{Vaidman, L.} \& \bibinfo{author}{Yoran, N.}
\newblock \bibinfo{title}{Methods for reliable teleportation}.
\newblock \emph{\bibinfo{journal}{Phys. Rev. A}} \textbf{\bibinfo{volume}{59,}}
  \bibinfo{pages}{116--125} (\bibinfo{year}{1999}).

\bibitem{lutkenhaus-pra-59-3295}
\bibinfo{author}{L\"{u}tkenhaus, N.}, \bibinfo{author}{Calsamiglia, J.} \&
  \bibinfo{author}{Suominen, K.~A.}
\newblock \bibinfo{title}{Bell measurements for teleportation}.
\newblock \emph{\bibinfo{journal}{Phys. Rev. A}} \textbf{\bibinfo{volume}{59,}}
  \bibinfo{pages}{3295--3300} (\bibinfo{year}{1999}).

\bibitem{kwiat-pra-58-R2623}
\bibinfo{author}{Kwiat, P.~G.} \& \bibinfo{author}{Weinfurter, H.}
\newblock \bibinfo{title}{Embedded {B}ell-state analysis}.
\newblock \emph{\bibinfo{journal}{Phys. Rev. A}} \textbf{\bibinfo{volume}{58,}}
  \bibinfo{pages}{R2623--R2626} (\bibinfo{year}{1998}).

\bibitem{schuck-prl-96-190501}
\bibinfo{author}{Schuck, C.}, \bibinfo{author}{Huber, G.},
  \bibinfo{author}{Kurtsiefer, C.} \& \bibinfo{author}{Weinfurter, H.}
\newblock \bibinfo{title}{Complete deterministic linear optics {B}ell state
  analysis}.
\newblock \emph{\bibinfo{journal}{Phys. Rev. Lett.}}
  \textbf{\bibinfo{volume}{96,}} \bibinfo{pages}{190501}
  (\bibinfo{year}{2006}).

\bibitem{barbieri-pra-75-042317}
\bibinfo{author}{Barbieri, M.}, \bibinfo{author}{Vallone, G.},
  \bibinfo{author}{Mataloni, P.} \& \bibinfo{author}{Martini, F.~D.}
\newblock \bibinfo{title}{Complete and deterministic discrimination of
  polarization {B}ell states assisted by momentum entanglement}.
\newblock \emph{\bibinfo{journal}{Phys. Rev. A}} \textbf{\bibinfo{volume}{75,}}
  \bibinfo{pages}{042317} (\bibinfo{year}{2007}).

\bibitem{molina-terriza-nphys-3-305}
\bibinfo{author}{Molina-Terriza, G.}, \bibinfo{author}{Torres, J.~P.} \&
  \bibinfo{author}{Torner, L.}
\newblock \bibinfo{title}{Twisted photons}.
\newblock \emph{\bibinfo{journal}{Nature Physics}}
  \textbf{\bibinfo{volume}{3,}} \bibinfo{pages}{305--310}
  (\bibinfo{year}{2007}).

\bibitem{barreiro-prl-95-260501}
\bibinfo{author}{Barreiro, J.~T.}, \bibinfo{author}{Langford, N.~K.},
  \bibinfo{author}{Peters, N.~A.} \& \bibinfo{author}{Kwiat, P.~G.}
\newblock \bibinfo{title}{Generation of hyperentangled photon pairs}.
\newblock \emph{\bibinfo{journal}{Phys. Rev. Lett.}}
  \textbf{\bibinfo{volume}{95,}} \bibinfo{pages}{260501}
  (\bibinfo{year}{2005}).

\bibitem{aolita-prl-98-100501}
\bibinfo{author}{Aolita, L.} \& \bibinfo{author}{Walborn, S.~P.}
\newblock \bibinfo{title}{Quantum communication without alignment using
  multiple-qubit single-photon states}.
\newblock \emph{\bibinfo{journal}{Phys. Rev. Lett.}}
  \textbf{\bibinfo{volume}{98,}} \bibinfo{pages}{100501}
  (\bibinfo{year}{2007}).

\bibitem{liu-pra-65-022304}
\bibinfo{author}{Liu, X.~S.}, \bibinfo{author}{Long, G.~L.},
  \bibinfo{author}{Tong, D.~M.} \& \bibinfo{author}{Li, F.}
\newblock \bibinfo{title}{General scheme for superdense coding between
  multiparties}.
\newblock \emph{\bibinfo{journal}{Phys. Rev. A}} \textbf{\bibinfo{volume}{65,}}
  \bibinfo{pages}{022304} (\bibinfo{year}{2002}).

\bibitem{harrow-prl-92-187901}
\bibinfo{author}{Harrow, A.}, \bibinfo{author}{Hayden, P.} \&
  \bibinfo{author}{Leung, D.}
\newblock \bibinfo{title}{Superdense coding of quantum states}.
\newblock \emph{\bibinfo{journal}{Phys. Rev. Lett.}}
  \textbf{\bibinfo{volume}{92,}} \bibinfo{pages}{187901}
  (\bibinfo{year}{2004}).

\bibitem{braunstein-pra-61-042302}
\bibinfo{author}{Braunstein, S.~L.} \& \bibinfo{author}{Kimble, H.~J.}
\newblock \bibinfo{title}{Dense coding for continuous variables}.
\newblock \emph{\bibinfo{journal}{Phys. Rev. A}} \textbf{\bibinfo{volume}{61,}}
  \bibinfo{pages}{042302} (\bibinfo{year}{2000}).

\bibitem{ban-qso-1-l9}
\bibinfo{author}{Ban, M.}
\newblock \bibinfo{title}{Quantum dense coding via a two-mode squeezed-vacuum
  state}.
\newblock \emph{\bibinfo{journal}{J. Opt. B: Quantum Semiclass. Opt.}}
  \textbf{\bibinfo{volume}{1,}} \bibinfo{pages}{L9} (\bibinfo{year}{1999}).

\bibitem{li-prl-88-047904}
\bibinfo{author}{Li, X.} \emph{et~al.}
\newblock \bibinfo{title}{Quantum dense coding exploiting a bright
  {E}instein-{P}odolsky-{R}osen beam}.
\newblock \emph{\bibinfo{journal}{Phys. Rev. Lett.}}
  \textbf{\bibinfo{volume}{88,}} \bibinfo{pages}{047904}
  (\bibinfo{year}{2002}).

\bibitem{fang-pra-61-022307}
\bibinfo{author}{Fang, X.}, \bibinfo{author}{Zhu, X.}, \bibinfo{author}{Feng,
  M.}, \bibinfo{author}{M, X.} \& \bibinfo{author}{Du, F.}
\newblock \bibinfo{title}{Experimental implementation of dense coding using
  nuclear magnetic resonance}.
\newblock \emph{\bibinfo{journal}{Phys. Rev. A}} \textbf{\bibinfo{volume}{61,}}
  \bibinfo{pages}{022307} (\bibinfo{year}{2000}).

\bibitem{schaetz-prl-93-040505}
\bibinfo{author}{Schaetz, T.} \emph{et~al.}
\newblock \bibinfo{title}{Quantum dense coding with atomic qubits}.
\newblock \emph{\bibinfo{journal}{Phys. Rev. Lett.}}
  \textbf{\bibinfo{volume}{93,}} \bibinfo{pages}{040505}
  (\bibinfo{year}{2004}).

\bibitem{kim-prl-86-1370}
\bibinfo{author}{Kim, Y.-H.}, \bibinfo{author}{Kulik, S.~P.} \&
  \bibinfo{author}{Shih, Y.}
\newblock \bibinfo{title}{Quantum teleportation of a polarization state with a
  complete {B}ell state measurement}.
\newblock \emph{\bibinfo{journal}{Phys. Rev. Lett.}}
  \textbf{\bibinfo{volume}{86,}} \bibinfo{pages}{1370--1373}
  (\bibinfo{year}{2001}).

\bibitem{calsamiglia-apb-B72-67}
\bibinfo{author}{Calsamiglia, J.} \& \bibinfo{author}{Lutkenhaus, N.}
\newblock \bibinfo{title}{Maximum efficiency of a linear-optical {B}ell-state
  analyzer}.
\newblock \emph{\bibinfo{journal}{Appl. Phys. B: Lasers Opt.}}
  \textbf{\bibinfo{volume}{B72,}} \bibinfo{pages}{67--71}
  (\bibinfo{year}{1999}).

\bibitem{kwiat-jmo-44-2173}
\bibinfo{author}{Kwiat, P.~G.}
\newblock \bibinfo{title}{Hyper-entangled states}.
\newblock \emph{\bibinfo{journal}{J. Mod. Opt.}} \textbf{\bibinfo{volume}{44,}}
  \bibinfo{pages}{2173--2184} (\bibinfo{year}{1997}).

\bibitem{allen-2003-oam}
\bibinfo{editor}{Allen, L.}, \bibinfo{editor}{Barnett, S.~M.} \&
  \bibinfo{editor}{Padgett, M.~J.} (eds.) \emph{\bibinfo{title}{Optical Angular
  Momentum}} (\bibinfo{publisher}{Institute of Physics Publishing},
  \bibinfo{address}{Bristol}, \bibinfo{year}{2003}).

\bibitem{walborn-pra-68-042313}
\bibinfo{author}{Walborn, S.~P.}, \bibinfo{author}{P\'{a}dua, S.} \&
  \bibinfo{author}{Monken, C.~H.}
\newblock \bibinfo{title}{Hyperentanglement-assisted {B}ell-state analysis}.
\newblock \emph{\bibinfo{journal}{Phys. Rev. A}} \textbf{\bibinfo{volume}{68,}}
  \bibinfo{pages}{042313} (\bibinfo{year}{2003}).

\bibitem{englert-pra-63-032303}
\bibinfo{author}{Englert, B.-G.}, \bibinfo{author}{Kurtsiefer, C.} \&
  \bibinfo{author}{Weinfurter, H.}
\newblock \bibinfo{title}{Universal unitary gate for single-photon two-qubit
  states}.
\newblock \emph{\bibinfo{journal}{Phys. Rev. A}} \textbf{\bibinfo{volume}{63,}}
  \bibinfo{pages}{032303} (\bibinfo{year}{2001}).

\bibitem{altepeter-oe-13-8951}
\bibinfo{author}{Altepeter, J.~B.}, \bibinfo{author}{Jeffrey, E.~R.} \&
  \bibinfo{author}{Kwiat, P.~G.}
\newblock \bibinfo{title}{Phase-compensated ultra-bright source of entangled
  photons}.
\newblock \emph{\bibinfo{journal}{Opt. Express}}
\textbf{\bibinfo{volume}{13,}}
  \bibinfo{pages}{8951--8959} (\bibinfo{year}{2005}).

\bibitem{wei-pra-2007}
\bibinfo{author}{Wei, T.-C.}, \bibinfo{author}{Barreiro, J.~T.} \&
  \bibinfo{author}{Kwiat, P.~G.}
\newblock \bibinfo{title}{Hyperentangled {B}ell-state analysis}.
\newblock \emph{\bibinfo{journal}{Phys. Rev. A}} \textbf{\bibinfo{volume}{75,}}
  \bibinfo{pages}{060305(R)} (\bibinfo{year}{2007}).

\bibitem{cerf-pra-57-R1477}
\bibinfo{author}{Cerf, N.~J.}, \bibinfo{author}{Adami, C.} \&
  \bibinfo{author}{Kwiat, P.~G.}
\newblock \bibinfo{title}{Optical simulation of quantum logic}.
\newblock \emph{\bibinfo{journal}{Phys. Rev. A}} \textbf{\bibinfo{volume}{57,}}
  \bibinfo{pages}{R1477--R1480} (\bibinfo{year}{1998}).

\bibitem{fiorentino-prl-93-070502}
\bibinfo{author}{Fiorentino, M.} \& \bibinfo{author}{Wong, F. N.~C.}
\newblock \bibinfo{title}{Deterministic controlled-not gate for single-photon
  two-qubit quantum logic}.
\newblock \emph{\bibinfo{journal}{Phys. Rev. Lett.}}
  \textbf{\bibinfo{volume}{93,}} \bibinfo{pages}{070502}
  (\bibinfo{year}{2004}).

\bibitem{ali-khan-prl-98-060503}
\bibinfo{author}{Ali-Khan, I.}, \bibinfo{author}{Broadbent, C.~J.} \&
  \bibinfo{author}{Howell, J.~C.}
\newblock \bibinfo{title}{Large-alphabet quantum key distribution using
  energy-time entangled bipartite states}.
\newblock \emph{\bibinfo{journal}{Phys. Rev. Lett.}}
  \textbf{\bibinfo{volume}{98,}} \bibinfo{pages}{060503}
  (\bibinfo{year}{2007}).

\bibitem{paterson-prl-94-153901}
\bibinfo{author}{Paterson, C.}
\newblock \bibinfo{title}{Atmospheric turbulence and orbital angular momentum
of single photons for optical communication}.
\newblock \emph{\bibinfo{journal}{Phys. Rev. Lett.}}
  \textbf{\bibinfo{volume}{94,}} \bibinfo{pages}{153901}
  (\bibinfo{year}{2005}).

\bibitem{smith-pra-74-062104}
\bibinfo{author}{Smith, B.~J.} \&
  \bibinfo{author}{Raymer, M.~G.}
\newblock \bibinfo{title}{Two-photon wave mechanics}.
\newblock \emph{\bibinfo{journal}{Phys. Rev. A}} \textbf{\bibinfo{volume}{74,}}
  \bibinfo{pages}{062104} (\bibinfo{year}{2006}).

\end{thebibliography}
\end{document}


\title{Beating the channel capacity limit for linear photonic superdense
  coding\\SUPPLEMENTARY INFORMATION}

\author{Julio T. Barreiro}
\author{Tzu-Chieh Wei}
\author{Paul G. Kwiat}
\affiliation{Department of Physics, University of Illinois at %
Urbana-Champaign, Urbana, Illinois 61801-3080, USA}


\maketitle 

\section{Supplement I: Channel capacity as a function of conditional
  detection average success probability (Figure 4b)}

In Fig. 4{\bf b}, the channel capacity upper bound for the 4-state encoding
domain occurs when two of the four states are faithfully transmitted and the
other two mixed, corresponding to a transfer matrix of conditional
probabilities $p(y|x)$ of receiving the state $y$ given that the state $x$
was sent,
\begin{equation}
\begin{pmatrix}
p(\Phi^+|\Phi^+) & p(\Phi^+|\Phi^-) & p(\Phi^+|\Psi^+) & p(\Phi^+|\Psi^-) \\
p(\Phi^-|\Phi^+) & p(\Phi^-|\Phi^-) & p(\Phi^-|\Psi^+) & p(\Phi^-|\Psi^-) \\
p(\Psi^+|\Phi^+) & p(\Psi^+|\Phi^-) & p(\Psi^+|\Psi^+) & p(\Psi^+|\Psi^-) \\
p(\Psi^-|\Phi^+) & p(\Psi^-|\Phi^-) & p(\Psi^-|\Psi^+) & p(\Psi^-|\Psi^-)
\end{pmatrix}
=
\begin{pmatrix}
2p_s-1 & 2(1-p_s) & 0 & 0\\
2(1-p_s) & 2p_s-1 & 0 & 0\\
0 & 0 & 1 & 0\\
0 & 0 & 0 & 1
\end{pmatrix}
\end{equation}
with an input distribution $p(x)$ maximized for each conditional detection
average success probability $p_s$.  The lower bound is set by a uniform noise
over the failed transmissions, resulting in a transfer matrix:
\begin{equation}
\begin{pmatrix}
p_s & f & f & f\\
f & p_s & f& f \\ 
f & f & p_s & f \\ 
f & f & f & p_s
\end{pmatrix}
,
\end{equation}
where $f=(1-p_s)/3$ and for all sent messages $x$, the input distribution
that maximizes the mutual information is $p(x)=1/4$.

\section{Supplement II: Effect of experimental imperfections on the channel capacity}

In our scheme, the major detriments to the channel capacity are the input
polarization and spatial-mode states and the PBS crosstalk; accidental
coincidences lead to only a minor \cc{} reduction.  We model each
imperfection as a function of relevant parameters, which were estimated by
quantum state tomography or measured.  Then we feed these parameters into a
Monte Carlo simulation of the expected channel capacity, as described below.

\subsection{Modelling imperfections: The input states}

The imperfections in the hyperentangled state generated by spontaneous
parametric downconversion,
$|\Phi^-_\text{spin}\rangle\otimes|\Psi^+_\text{orbit}\rangle$ in our case,
can be modelled with high fidelity by uniform-noise decoherence, a population
imbalance angle $\epsilon_\theta$ and a relative phase $\epsilon_\phi$.  The
deviations from the ideal pure state
$|\Phi^-_\text{spin}\rangle\otimes|\Psi^+_\text{orbit}\rangle$ have the
following form

\begin{eqnarray}
|\Phi^-_\text{spin}(\epsilon_{\theta,\text{spin}},\epsilon_{\phi,\text{spin}})\rangle&=&
\cos(\pi/4+\epsilon_{\theta,\text{spin}})|HH\rangle -
e^{i\epsilon_{\phi,\text{spin}}}\sin(\pi/4+\epsilon_{\theta,\text{spin}})
|VV\rangle,\label{eq:phispin}\\ 
|\Psi^+_\text{orbit}(\epsilon_{\theta,\text{orbit}},\epsilon_{\phi,\text{orbit}})\rangle
&=& \cos(\pi/4+\epsilon_{\theta,\text{orbit}})|\oaml\oamr\rangle+ e^{i
  \epsilon_{\phi,\text{orbit}}}\sin(\pi/4+\epsilon_{\theta,\text{orbit}})|\oamr\oaml\rangle.\label{eq:psiorbit}
\end{eqnarray}

The uniform-noise decoherence is characterized by a parameter $\lambda$ and
it adds mixture to each spin and orbital density matrices in the form
$\rho(\lambda) = (1-\lambda)\rho+\lambda\mathbf{I}_4$, where $\mathbf{I}_4$
is the 4-dimensional identity matrix.  Since the imperfections also affect
the messages encoded, the four states in the dense-coding protocol have a
functional dependence on the source parameters $\epsilon_\theta$,
$\epsilon_\phi$ and $\lambda$:

\begin{equation}
\begin{split}
\Phi^\pm_\text{spin}(\epsilon_{\theta,\text{spin}},\epsilon_{\phi,\text{spin}},\lambda_\text{spin})\otimes
\Psi^+_\text{orbit}(\epsilon_{\theta,\text{orbit}},\epsilon_{\phi,\text{orbit}},\lambda_\text{orbit})\\
\Psi^\pm_\text{spin}(\epsilon_{\theta,\text{spin}},\epsilon_{\phi,\text{spin}},\lambda_\text{spin})\otimes
\Psi^+_\text{orbit}(\epsilon_{\theta,\text{orbit}},\epsilon_{\phi,\text{orbit}},\lambda_\text{orbit})
\end{split}
\end{equation}

Moreover, since these parameters vary slightly due to alignment~\cite{mynote}
and component imperfections among the different combinations of signature
detectors, we performed a quantum state tomography in each combination and
determined the spread in the parameters.  To estimate the channel capacity,
we performed Monte Carlo simulations where the above parameters are the
random variables following a normal distribution with a standard deviation
given by the spread of the estimated parameters.

\subsection{Modelling imperfections: The spin-orbit gate}

The forked hologram maps the \oam{} states $|\oaml\rangle$ and
$|\oamr\rangle$ into distinct propagation modes $a$ and $b$ with no \oam,
respectively corresponding to the $+1$ and $-1$ diffraction orders (with a
$\pi$ phase shift difference):

\begin{equation}
\left(
\begin{array}{c}
  |H\oaml\rangle\\
  |H\oamr\rangle\\
  |V\oaml\rangle\\
  |V\oamr\rangle
\end{array}
\right)
\xmapsto{\rm \;\;forked\;\;hologram\;\;}
\begin{pmatrix}
  |H_a\rangle\\
  -|H_b\rangle\\
  |V_a\rangle\\
  -|V_b\rangle
\end{pmatrix}.
\label{eq:holo}
\end{equation}

The polarizer beam splitter (PBS) is modelled with a crosstalk parameter for
each polarization, $\epsilon_H$ and $\epsilon_V$, and a relative phase shift
between $H$ and $V$ polarizations for each output mode, $\phi_1$ and
$\phi_2$.  In the basis of Eq.~\ref{eq:holo}, 

\begin{equation}
PBS(\epsilon_H,\epsilon_V,\phi_1,\phi_2)=
\left(\begin{array}{cccc}
 \sqrt{1-\epsilon_H} & -\sqrt{\epsilon_H} & 0 & 0 \\
 \sqrt{\epsilon_H} & \sqrt{1-\epsilon_H} & 0 & 0 \\
 0 & 0 & e^{\frac{1}{2} i (\phi_1+\phi_2)}
 \sqrt{\epsilon v} & -e^{i \phi_2} \sqrt{1-\epsilon
     v}
   \\
 0 & 0 & e^{i \phi_1} \sqrt{1-\epsilon v} & e^{\frac{1}{2}
   i (\phi_1+\phi_2)} \sqrt{\epsilon v}
\end{array}\right).
\end{equation}
The phase shifts $\phi_1$ and $\phi_2$ were corrected with tilted
birefringent plates (quarter-wave plates, in particular) with their optic
axes aligned parallel to the direction of horizontal polarization.
Therefore, for the model we can set $\phi_1=\phi_2=0$, and the PBS depends
only on the crosstalk, $PBS(\epsilon_H,\epsilon_V)$.

The gate for the pair of photons is therefore
\begin{equation}
U(\epsilon_H,\epsilon_V) = \left(PBS(\epsilon_H,\epsilon_V)\cdot
\text{Hologram}\right)_\text{photon
  1}\otimes\left(PBS(\epsilon_H,\epsilon_V)\cdot
\text{Hologram}\right)_\text{photon 2}.
\end{equation}

\subsection{Estimating the channel capacity}

Using the above input states and spin-orbit gates we can estimate the
conditional detection probabilities; for example, the probability of
inferring $\Phi^+$ given that $\Phi^-$ was sent is
\begin{equation}
p(\Phi^+|\Phi^-) = \langle\Phi^+_\text{spin}\otimes\Psi^+_\text{orbit}|
U(\epsilon_H,\epsilon_V) \rho_{\Phi^-_\text{spin}\otimes\Psi^+_\text{orbit}}
U(\epsilon_H,\epsilon_V)^\dagger|\Phi^+_\text{spin}\otimes\Psi^+_\text{orbit}\rangle,
\end{equation}
where $\rho_{\Phi^-_\text{spin}\otimes\Psi^+_\text{orbit}}$ is a function of
$\epsilon_{\theta,\text{spin}}$, $\epsilon_{\phi,\text{spin}}$,
$\lambda_\text{spin}$, $\epsilon_{\theta,\text{orbit}}$,
$\epsilon_{\phi,\text{orbit}}$ and $\lambda_\text{orbit}$.  Finally, the
channel capacity is obtained by maximizing the mutual information over all
the input distributions.

\subsection{Results}

We characterized the source polarization state $\Phi^-_\text{spin}$ by
quantum state tomography in the $|\oaml\oamr\rangle$ and $|\oamr\oamr\rangle$
\oam{} subspaces~\cite{barreiro-prl-95-260501}.  Considering all combinations
of signature detectors, we measured an average degree of entanglement or
tangle of $T=96.7(8)\%$ and a mixture or linear entropy of $S_L=2.0(4)\%$.
The \oam{} state was also tomographically reconstructed in the $|HH\rangle$
and $|VV\rangle$ polarization subspaces~\cite{barreiro-prl-95-260501},
measuring an average $T=91(3)\%$ and $S_L=6(2)\%$.  We calculated the states
of the form of Eq.~\ref{eq:phispin} and~\ref{eq:psiorbit} closest to the
reconstructed density matrices.  The fidelity between the reconstructed and
closest estimated states is 99\% for polarization and 97\% for spatial-mode
states.  The estimated parameters and their spread are shown in the table
below.  The achievable channel capacity under only one imperfection is quoted
as well.  The Monte Carlo simulations used 100 iterations, enough to display
stable values in the first digit of the uncertainty.

\begin{table}
\begin{tabular}{|c|c|c|c|}\hline
\multirow{2}{*}{Imperfection} & Conditional detection &
Achievable \cc{} & \cc{} reduction\\
& average success probability & (bits) & (bits)\\\hline
Source polarization state&\multirow{3}{*}{0.991(2)}&\multirow{3}{*}{1.91(2)}&\multirow{3}{*}{0.09(2)}\\
$\epsilon_{\theta,\text{spin}}=1.0(7)^\circ$,
$\epsilon_{\phi,\text{spin}}=0(4)^\circ$,
$\lambda_\text{spin}=0.010(2)$&&&\\
$T=96.7(8)\%$ , $S_L=2.0(4)\%$&&&\\\hline
Source spatial-mode state &\multirow{3}{*}{0.971(4)}&\multirow{3}{*}{1.80(3)}&\multirow{3}{*}{0.20(3)}\\
$\epsilon_{\theta,\text{orbit}}=1.7(6)^\circ$,
$\epsilon_{\phi,\text{orbit}}=0(5)^\circ$,
$\lambda_\text{orbit}=0.03(1)$&&&\\
$T=91(3)\%$, $S_L=6(2)\%$&&&\\\hline
PBS crosstalk &\multirow{3}{*}{0.985(2)}&\multirow{2}{*}{1.90(1)}&\multirow{2}{*}{0.10(2)}\\
$\epsilon_H=0.005(1)$, $\epsilon_V=0.010(2)$&&&\\\hline
Accidentals (5 in 150s) & 0.998(1) & 1.98(1)&0.02(1)\\\hline
All the above imperfections in the same simulation & 0.953(3) & 1.64(2) & 0.36(2)\\\hline
\end{tabular}
\end{table}

The subtle and convoluted character of the imperfections is evident upon
directly adding the independently modeled CC reductions of each imperfection,
which would predict an anomalously low CC=1.59 bits, compared to the full
Monte-Carlo simulation resulting from simultaneously modelling all the
effects, which predicts CC=1.64(2) bits.

Finally, we stress that the reported conditional distribution was not
normalized by the statistical detection probabilities to account for
detection imbalances over the different states; these were well balanced in
our case, only 3\% off for the states $\Phi^-$ and $\Psi^+$, such
normalization would increase our reported CC to 1.634(6) bits.

\begin{figure}
\centerline{\includegraphics[width=15cm]{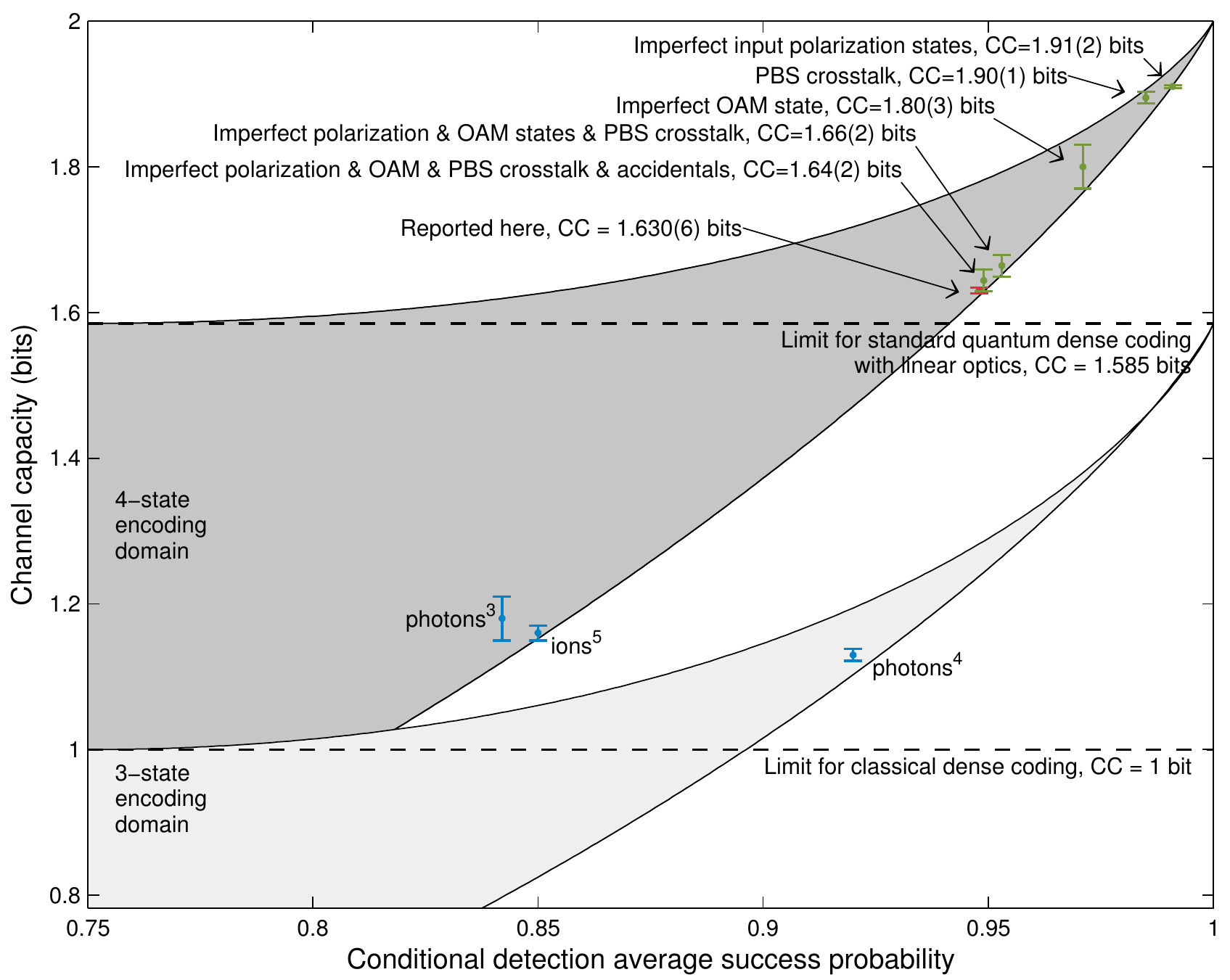}}
\caption{\textbf{Experimentally achieved and simulated channel capacities as
    a function of conditional detection average success probability.}  The
  simulated \cc{} for our hyperentanglement-assisted dense-coding protocol,
  when affected by only the experimental imperfections indicated, are shown
  in green.  The \cc{} reported in the accompanying article is shown in red.
  Previously reported \cc{}s are shown in blue for photons
  with~\cite{schuck-prl-96-190501} and without
  hyperentanglement~\cite{mattle-prl-76-4656} and for
  ions~\cite{schaetz-prl-93-040505}.  }
\label{fig:counts}
\end{figure}
